# Scoping Studies for NBI Launch Geometries on DEMO

I. Jenkins, C.D. Challis, D.L. Keeling, E. Surrey

*EURATOM/CCFE Fusion Association, Culham Science Centre, Abingdon, Oxon OX14 3DB, UK*

**Abstract**

Scans of Neutral Beam Injection (NBI) tangency radii and elevation on two possible DEMO scenarios have been performed with two beam energies, 1.5MeV and 1.0MeV, in order to determine the most favourable options for Neutral Beam Current Drive (NBCD) efficiency. In addition, a method using a genetic algorithm has been used to seek optimised solutions of NBI source locations and powers to synthesize a target total plasma driven-current profile. It is found that certain beam trajectories may be proscribed by limitations on shinethrough onto the vessel wall. This may affect the ability of NBCD to extend the duration of a pulse in a scenario where it must complement the induced plasma current. Operating at the lower beam energy reduces the restrictions due to shinethrough and is attractive for technical reasons, but in the scenarios examined here this results in a spatial broadening of the NBCD profile, which may make it more challenging to achieve desired total driven-current profiles.

# 1. Introduction

At the conceptual design phase of a proposed device such as DEMO, the definition of the design and R&D needs for Heating and Current Drive (H&CD) systems are strongly linked with the assumed requirements. These requirements can be strongly influenced by plasma parameters. In the case of Neutral Beam Injection (NBI) systems the kinetic profiles can have a critical effect on beam deposition profiles, as well as resulting quantities such as Neutral Beam Current Drive (NBCD), determining the feasibility of injecting a beam at a given energy.

Given the large range of possibilities for system configurations at the conceptual design phase, a time efficient method of scanning through possible options is required in order to narrow down the ranges of, for example, NBI tangency radii and elevations at different beam energies. Such a method must be adaptable to each DEMO design and scenario as differing balances of heating and current drive have been proposed.

In this study a range of possible NBI trajectories are scanned, allowing a contour map of NBCD efficiency to be generated for a given set of plasma density and temperature parameters, corresponding to possible DEMO scenarios. These scans also generate an NBCD profile for each source location which can be used in an iterative process to find possible NBI geometries and power levels that, for given plasma kinetic profiles, will synthesize a required NBCD target profile.

Such a target profile can be derived from the difference between the desired total current profile and the bootstrap current profile from a suitable discharge, thus simulating the NBCD required for partial, or even total, replacement of the ohmic current in envisaged current drive-assisted enhanced pulse or steady-state operation scenarios.

The code predominantly used in these studies is PENCIL [1]. This models a beam with many parallel non-divergent beamlets, each having a weighting resulting in a Gaussian power distribution over the beam envelope. Beam-stopping coefficients are from the ADAS database [2], and these have been extended to the high beam energies used.

The beam particle deposition is calculated at each point along the trajectories of the beamlets within the plasma. This gives a source function for a Fokker-Planck code to model the fast ion energy distribution in the plasma allowing a comparatively rapid run time compared to a Monte Carlo approach, advantageous when undertaking parameter space scans. Gyro-orbit effects are not included but these are not seen as important in DEMO as the normalized gyro-radius is small in comparison with the plasma major radius.

Certain analytic approximations are used in the code, and these have been subject to revision and improvement e.g. the calculation of screening by trapped electrons on the driven current has recently been improved by introducing a model that accounts for the plasma collisionality varying across the plasma profile [3].

## 2. Parameter Scans for NBCD on DEMO with $E_{beam}$=1.5MeV & 1.0MeV

Two sets of kinetic profiles were used in these scans, as shown in Fig.1, and are referred to as 'flat' and 'peaked' after the characteristic difference in the shape of the density profile. The temperature data describe both ion ($T_i$) and electron ($T_e$) temperatures, assumed equal, in these simulations and the profiles correspond to two possible DEMO scenarios. A Z-effective of 1.57 was assumed uniformly across the plasma, the sole impurity being Argon, purposely used with the intention of radiating power to reduce the power loading on the divertor.

For a given DEMO plasma geometry, up-down symmetric about the plasma axis, with major radius, R=9.18m, minor radius, a=2.3m, elongation, ε~1.5, triangularity, δ~0.4, the effect of NBI for varying tangency radius, $R_T$, and elevation Z was investigated to reveal the influence on NBCD magnitude and efficiency. The values of $R_T$ varied from 7.4m to 10.6m in 0.2m increments and Z from 0 (plasma axis) to 3m, in 0.2m increments; beam energy, $E_B$ was 1.5MeV.

As a point of comparison, TRANSP [4] runs were also made for a single tangency of 9m, Z=0 with a 1MW, 1.5MeV beam for both scenarios to compare with the PENCIL output for that location. The total NBCD agreed on both occasions to within a few per cent.

Resulting contours of NBCD efficiency, γ, can be plotted from the PENCIL scan output, calculated using $\gamma = (n_e\ I_{CD}\ R_T\ )/P_{inj}$ with $I_{CD}$ (MA) representing the total driven current in the plasma at tangency radius $R_T$ (m) and injected power $P_{inj}$ (MW). The density term, $n_e$ is strictly the volume averaged density along the beam trajectory normalised to $1\times10^{20}$ m$^{-3}$, but usually the volume averaged density $<n_e>$ over the whole plasma is substituted as many differing beam trajectories are involved. The value of $<n_e>$ is calculated by PENCIL as $1.1\times10^{20}$ m$^{-3}$ for the peaked density profile and $8.9\times10^{19}$ m$^{-3}$ for the flat density profile.

The resulting values of γ at all injection trajectories in the scan can be amalgamated into contour maps to give a projection of γ at the vertical mid-plane over the plasma. For the flat and peaked density profiles respectively, these are given in Fig.2. It can be seen that higher values of γ are attained in the flat density scenario, where the lower values of density and the higher temperature both favour NBCD, compared to the peaked scenario. (It should be noted that the values of γ presented here differ from those presented in a previous study [5] as the reference scenarios have changed with regards to many parameters, including the size of the plasma and, in particular, the plasma temperature.)

However, the lower line-integrated density encountered by the beam in the flat density profile will leave a greater proportion of the beam un-ionized and striking the tokamak inner wall, a phenomenon known as 'shinethrough'. In evaluating possible options for DEMO injectors and beamlines in [5], it was estimated that the peak power density for a 1.5MeV beam entering the vessel is of order 300MW/m$^2$. As the nominal allowable wall power loading is set at 2MW/m$^2$ [6] for all sources of power, setting a limit of 0.75MW/m$^2$ for heating due to the NBI alone means that a lower shinethrough limit should be set at 0.25%. Contours of percentage shinethrough showing the location of the 0.25% limit are shown for both scenarios in Fig. 3. This would exclude, for instance, injection geometries with $R_T$ > 9.6m for Z=0 for the flat density case.

The situation is slightly less restrictive for the peaked density case where a 0.25% limit allows injection up to $R_T$=10.2m at Z=0. Although this still allows injection at values of γ=0.5 and γ=0.4 for the flat and peaked density profiles respectively, the implications of such a restriction on the injection tangency become clear when considering the resulting spatial distribution of the NBCD profile (see section 4).

One way of visualizing this is to locate the maximum of each of the NBCD profiles from the scan in $R_T$ and Z. The position of the maximum can be plotted on the scale of the plasma minor radius, *r/a*, where *r* is the distance from the magnetic axis to the peak of the NBCD profile and *a*, the plasma minor radius with *r/a*=0 being the magnetic axis and *r/a*=1 the edge. The position of the maxima for the entire scan can then be plotted as contours as shown in Fig.4. For the injection geometries which do not pass close to the magnetic axis, the maximum of the NBCD profile will be located nearer to the plasma edge than for those which do approach the axis at some point in their trajectory. For example, injection at R=10.6m, Z=0 into the flat density profile, as shown in Fig.4a, causes an NBCD profile to be peaked towards the edge of the plasma, the maximum occurring at *r/a*=0.8. Injecting at R=9.5m, Z=0 causes the NBCD to be peaked closer to the magnetic axis at *r/a*=0.2. However, injecting inboard of the magnetic axis at Z=0 means that the beam will still pass through the magnetic axis at some point in its trajectory due to toroidal geometry. The resulting fast particle deposition from such a trajectory will lead to the maximum of the NBCD still being located close to the magnetic axis.

The position of these maxima can then be plotted as contours as shown in Fig.4 with, for example, the contour '0.4' showing the locus of injection tangencies and elevations where the peak NBCD will be at $r/a$=0.4. These contours give an indication as to whether current will be driven on-axis (for $r/a \leq 0.4$) or off-axis ($r/a \geq 0.6$). The 0.25% shinethrough limit is also indicated and it can be seen that this limit restricts where off-axis NBCD could be injected, particularly for the flat density case where injection at Z >1.5m, $R_T$ <9m would be necessary not to exceed it.

There are various technical issues surrounding NBI operation at high voltages which would make the possibility of operating at $E_B$=1MeV an attractive alternative to 1.5MeV if the physics requirements could be achieved. This would enable the NBI systems developed for ITER to be adapted, with some modifications, thus minimising the technological advances required for DEMO.

Repeating the scans for $E_B$=1 MeV gives resulting values of γ as shown in Fig. 5. Comparison with the $E_B$=1.5MeV scan shows that the values of γ for any given tangency and elevation are not reduced significantly for either the flat or peaked density scenarios. As the gamma value pertains to the total value of the driven current at that specific injection tangency and elevation, this does not give any indication of how well specific target current profiles can be driven at the lower beam energy. This will be discussed in section 4.

The reduction in beam energy to 1 MeV will reduce the peak power density in the vessel to 200MW/m$^2$ which, for the same 0.75MW/m$^2$ limit for NBI on the inner wall, will lead to a shinethrough limit of 0.375%. The position of this limit for 1 MeV beams in these scenarios is shown in Fig. 6. This shows that the options are increased at the lower beam energy, allowing injection tangencies up to ~10.3m at Z=0 for the flat density case, nearly as far as the ~10.4m at Z=0 allowed for the peaked one. As can be seen from the positions of the NBCD maxima in Fig.7 (the analogue of figure 4), off-axis NBCD can now be far more easily accessed in the flat density scenario. This ability to drive current off-axis may become important for enhanced pulse scenarios where NBCD complements the ohmic current (see section 5).

## 3. Use of Genetic Algorithm to Optimise NBI Source Placement.

The scans generate an array of NBCD source functions, which for each of the source positions in the scan shows the contribution to the total current drive per unit power injected. Thus, it is possible to synthesize a given target NBCD profile from a combination of these source functions, indicating which source positions would be required, along with their estimated powers. For reasons of economy and minimising interruption to the breeding blanket, it is advantageous to operate with the minimum number of beamlines of minimum acceptable power. Obtaining the optimal source array that satisfies these requirements in addition to that of the CD profile is achieved by the use of a genetic algorithm (GA).

The main problem is the size of the dataset as the number of injector positions tested with the PENCIL code occupies a grid with 17×17 positions. Furthermore, if each injector is assumed to have a power between 0 and 255MW with a resolution of 1MW then the number of possible configurations is given by $n^m$ where n is the number of positions (17x17) and m is the number of power levels each injector may have, i.e. $(17x17)^{256}$ or $10^{630}$ possible configurations. Obviously it is not possible to test every possible configuration, and even testing a subset of the possible configurations with a lower spatial resolution and lower power resolution still presents too large a search space to be a computationally tractable problem.

Genetic search algorithms (GAs) [7-9] have been successfully used in a wide variety of optimisation problems where a solution is required from a very large number of possible combinations. Such an algorithm has been developed which is capable of finding solutions to the many-valued problem of multiple plasma scenarios with multiple target current-drive profiles though only a single scenario, single target will be discussed here.

In general, a GA attempts to evolve a population of trial solutions towards the global optimum solution. In this sense, a trial solution is a binary encoded representation of a possible solution to the problem under investigation that may be easily manipulated to produce variations. The quality of each trial solution is assessed using a "fitness function" which is designed to tend to a maximum value as the solution tends toward the global optimum; the value of the function for each trial solution is referred to as the "fitness" of the solution. In each generation, the fitness of each trial solution is calculated then solutions are selected in proportion to their fitness values for inclusion in the next generation. The process of "crossover" is then applied to the selected solutions in the new generation to generate new trial solutions. In the simplest implementation crossover involves randomly selecting a position in the binary encoded bit string and exchanging all the bits after that point between the two trial solutions. In so doing, a large number of new trial solutions are formed some of which, by including the best traits from the previous generation, will have improved fitness scores. (Many will also have worse fitness scores but will therefore have a reduced probability of being selected to continue into later generations.) "Mutation" may then be applied by which a few randomly selected bits in the trial solutions are flipped (i.e. the selected bit, if 0, becomes 1 and vice-versa), which maintains diversity amongst the population and helps prevent convergence on sub-optimal solutions (though this is not guaranteed).

The Crossover and Mutation processes must be set to occur at certain rates in order to achieve a sufficiently rapid evolution of the population towards a global solution but not so rapid that convergence on local, sub-optimal solutions is favoured. If fixed crossover and mutation rates are applied globally throughout the execution of the G.A. run, selection of the particular values they should take becomes a problem. There is as yet no generally agreed method for selecting the values for these rates other than by executing a number of test runs to determine how the GA responds. Another method, employed in the present study, is to use variable rates that change as the population evolves. The particular scheme used is as described in [10] in which the rates for crossover/mutation are specific to each trial solution and encoded within the solutions themselves. Thus the crossover and mutation processes that evolve the solutions also evolve the rates at which these processes occur.

During tests, it was found that this scheme gave far superior performance compared with using fixed global rates in terms of evolving to more optimal solutions in fewer generations. Another possible issue that can cause GAs to prematurely converge on sub-optimal solutions is the approach for selection of solutions for progression to the next generation. Simple proportional selection (sometimes known as "roulette" selection), can introduce a very high degree of selection pressure that forces premature convergence on sub-optimal solutions.

Various methods for reducing the probability of premature convergence are discussed widely in the literature and some of these have been implemented, particularly the use of "Tournament selection" and the forced inclusion of solutions with lower fitness values. These techniques are well documented in the literature so will not be described in detail. The use of these techniques ensures that the global search-space is sufficiently sampled to lend confidence that the eventual solution is a sufficiently good representation of the global optimum.

The trial solutions used in this study each consist of two bit-strings. String 1 represents the source function selection and is 1 bit-per-source, indicating the presence or absence of a particular source in the solution. String 2 represents the power scaling factor for each source and is 8 bits-per-source giving a power resolution of 256 equally spaced levels between 0 and the maximum allowed power-per-source. Each bit string also includes a binary representation of the crossover and mutation rates associated with that particular trial solution and bit-string. A schematic representation of such a trial solution is shown in Fig 8. In the crossover process, the bit strings from one trial solution exchange information with the bit-strings from another trial solution, however information is only exchanged between same numbered bit-strings. This separation allows for simple adaptation of the problem to multi-scenario/multi-target cases by addition of further bit strings to each trial solution, with each bit-string representing a different combination of power levels to be used for the solution of a different scenario/target combination whilst maintaining the same source selection.

This explicitly encodes the idea that a beam system will have a fixed geometry but the power levels of the installed beam may be varied between zero and the maximum. Tests were carried out on such a problem in which 3 different target current-drive profiles were required in two different plasma scenarios giving 6 target/scenario combinations in total. Whilst the final solution was not as optimized as could be obtained for a single scenario/single target problem, a satisfactory solution was found and the computational cost per generation only increased approximately linearly with the number of target/scenario combinations.

The fitness function used is based on the least-squares-difference between the synthesized current profile and the target. The basic fitness is then defined as

$$f_{basic} = \sqrt{\overline{(target^2)}} - \sqrt{\overline{[(trial\ solution - target)^2]}}$$

This function approaches its maximum value in an approximately linear fashion as the solution approaches the optimum. For multi-scenario/multi-target problems, a weak exponential term could be introduced to prevent favouring particular combinations in the solution over others. Two modifiers to the fitness are then applied to allow optimization for solutions which contain fewer individual sources and lower total power, both of which are desirable to reduce the cost of the final beam-system.

The modifiers were constructed so as to reduce fitness scores for trial solutions with higher power and more selected sources and are simply introduced as multiplicative factors to the basic fitness

$$f_{total} = \left[\frac{\sum_{n=1}^{num.combinations}(f_{basic,n})}{num.\ combinations}\right] \times Mod_{num\ sources} \times Mod_{power}$$

Thus, in cases where these modifiers are used, the optimum solution may not have quite as good a fit to the target current drive profile(s) but will achieve a desirable compromise between fitting to the target and the eventual cost of the beam system. Although only two have been used, further modifiers can be included in the fitness function to represent other aspects of the beam system and weighting factors can be included (such as cost per unit power) to indicate the relative importance of each aspect.

## 4. Application of GA to target j-profile matching.

The target profile (Fig. 9) is derived from a 'hybrid' type q-profile with q>1 everywhere and flat q in the core typical of such used on JET [11], seen as a promising choice for DEMO. Although it is not intended to drive all the current by beams in a near-term DEMO, estimating the power required to do so could prove instructive as it would help show the relative impact on NBCD power demands of the differing plasma scenarios, and also give an indication of how closely a desired current profile could be matched. TRANSP simulations were run with the two sets of kinetic profiles and the given q-profile in order to calculate the resulting bootstrap currents. The difference between the total current profile and the bootstrap current profile in each case gives the current profile that must be provided by the summation of all inductive and non-inductive current drives. This gives a resultant current requirement of 10.37MA (14.05MA total current - 3.69MA bootstrap current) for the flat density case and 9.28MA for the peaked density (14.05MA - 4.77MA, here the bootstrap current is higher). Though it is not proposed to totally replace the ohmic component of the plasma current with non-inductive current drive on DEMO, as done in these fits, this method could also be used to fit a proposed NBCD fraction of the total current to complement ohmic current drive (i.e. 'enhanced pulse') when the details of a proposed mix are decided.

Scans were again done for 1.5MeV and 1MeV beam energies in order to generate NBCD profiles for each injection trajectory. For this exercise, the beam trajectories were inclined so that they all crossed the mid-plane at R=15m (a possible location for an NBI duct on DEMO) covering an angular range of 0 to 12 degrees, equivalent to a range of Z from 0 to ~2.8m at the beam tangency point. The genetic algorithm code was run for 1000 generations and resulting fits for the $E_B$=1.5MeV NBCD source functions to the target j-profile derived from the flat and peaked density profiles respectively, are shown in Fig.10. These have been optimised to minimise the number of individual sources required. It can be seen that a considerably higher injected power is required to achieve the best fit for the peaked density case relative to the flat density (474MW vs. 250MW, though, as stated, this level of power is not expected to be used in a near-term DEMO). This is due to the requirement to drive current predominantly on-axis (i.e. r/a<0.4 – see Fig.4b) which, for the peaked density case, causes the sources to be grouped so that the resulting beam tangency radii are towards the innermost point of the scan ($R_T$=7.4m) where the NBCD efficiency, γ, is lower for the peaked density scenario than for the flat one (0.18 vs. 0.3 - see Fig. 2). In fact, at this innermost position the driven currents differ by a factor of ~2 (20kA/MW for peaked vs. 38kA/MW for flat at

R=7.4m, Z=0).The fit for the flat density case gives a driven current of 10.89MA (target 10.37MA) and the peaked case fit gives a driven current of 9.74MA (target 9.28MA).

The matching fits for the $E_B$=1.0MeV case are shown in Fig.11 where it can be seen that considerably greater power is required at the lower beam energy in both scenarios to drive the current required for the best fit to the target j-profile. It can be seen from Fig. 11 that the best fits for the 1 MeV cases are not as good as with the 1.5MeV beam, reflected in the higher values of Root Mean Square Difference (RMSD) between the achieved profile and the target, even if, in the case of the flat profile (Fig. 11a), the best fit is achieved with one fewer source location at 1MeV compared to 1.5MeV (5 as opposed to 6). This is linked with the NBCD profiles having a broader spatial distribution within the plasma at the lower beam energy resulting in more edge current being driven. This is especially apparent in the peaked density profile where, despite all the power being from sources with $R_T$ =7.4m where the most on-axis current is driven, it can be seen that substantial current is being driven at the edge. For the flat density case the driven current is 11.00MA (target 10.37MA), the area integral effect of the off-axis current results in a driven current of 10.76MA for the peaked density case, considerably more than the target value of 9.28MA.

**5. Conclusions.**

The launch geometry scans for $E_B$=1.5 MeV (Figs 2-4) show that the NBCD efficiency is higher for the flat density case than for the peaked. However, the shinethrough limit at this beam energy may restrict the allowed options in the flat density plasma for off-axis current drive. This may become important in enhanced pulse scenarios where the beams are driving part of the plasma current in conjunction with ohmic current drive as the induced drive will tend to relax to a current profile which is peaked on-axis. If a hybrid-like current profile were desired, the NBCD would be required to drive current off-axis to complement the ohmic current and maintain a combined current profile more like that shown in Fig. 9.

This issue is reduced in importance if $E_B$ is reduced to 1 MeV as the degree of flexibility for off-axis drive is increased for both flat and peaked cases also the choice of 1MeV may become desirable for technical reasons. It should, however, be noted that the broader spatial distribution of the NBCD profile at 1MeV may affect the size of the area of the plasma where q is relatively flat ('low shear' region) in the hybrid scenario.

The use of genetic algorithms can help in seeking optimized solutions in large multi-variant spaces such as the problem of finding best source locations/powers to provide a given NBCD profile, as illustrated here. Although, as stated, it is not intended that the entire current profile be driven by NBCD in a near-term DEMO as the power levels required would (as shown here) be extremely large, this method will be useful for fitting the NBCD component of a compound current profile which would also contain ohmic and bootstrap current components, consistent with proposed enhanced pulse scenarios. Although this method of fitting a current profile output from a TRANSP simulation is not self-consistent in that it does not model the effect of the heating power on the equilibrium, this method, by showing optimal locations for NBI sources to provide desired NBCD results in given plasma scenarios, can help provide requirements for beamline geometry on DEMO.

**Acknowledgement**

This work was funded by the RCUK Energy Programme [under grant EP/I501045] and the European Communities under the contract of Association between EURATOM and CCFE. This work was carried out within the framework of the European Fusion Development Agreement. To obtain further information on the data and models underlying this paper please contact PublicationsManager@ccfe.ac.uk. The views and opinions expressed herein do not necessarily reflect those of the European Commission.

**Figures**

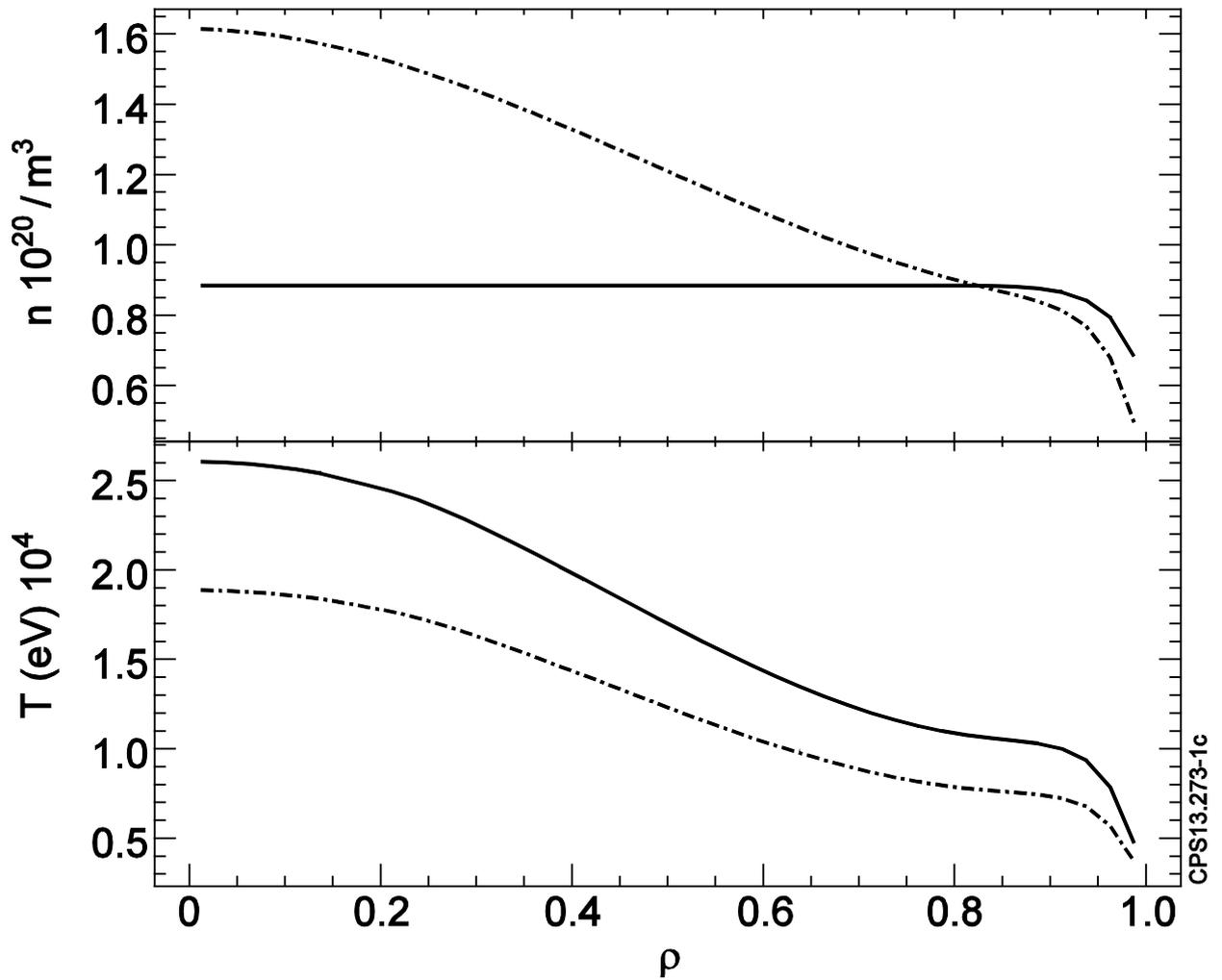

Fig. 1. Density (upper) and Temperature (lower, Ti=Te) profiles plotted against rho for the DEMO scenarios considered. The solid lines relate to the 'flat' density profile case and the dashed to the 'peaked' density profile case.

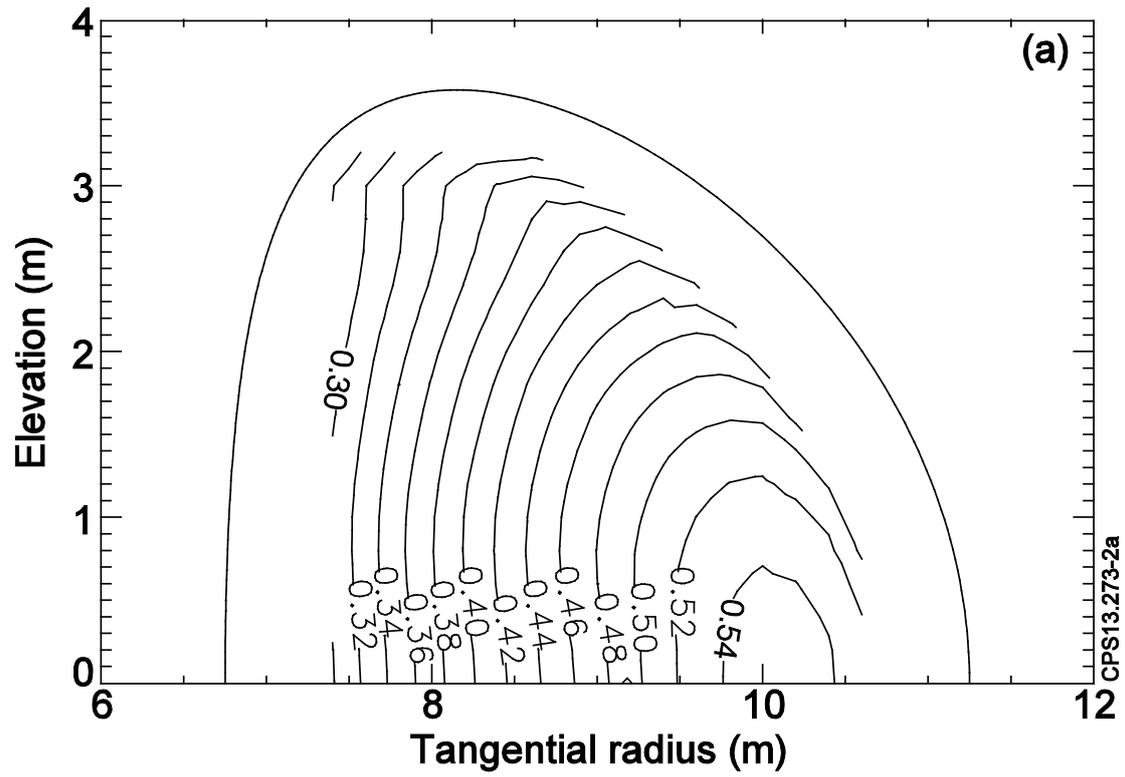
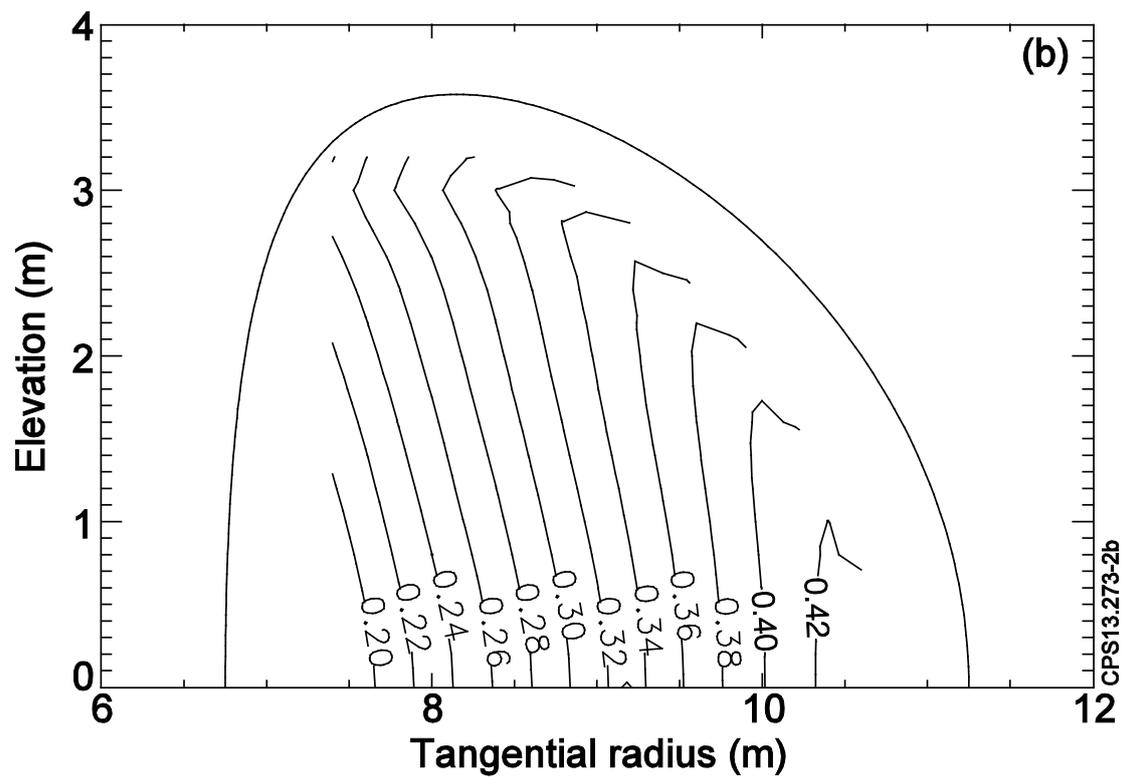

Fig. 2. Contours of NBCD efficiency, $\gamma$, with $E_B=1.5\text{MeV}$ for a range of tangency radii and elevations for a) the flat density scenario, b) the peaked density scenario.

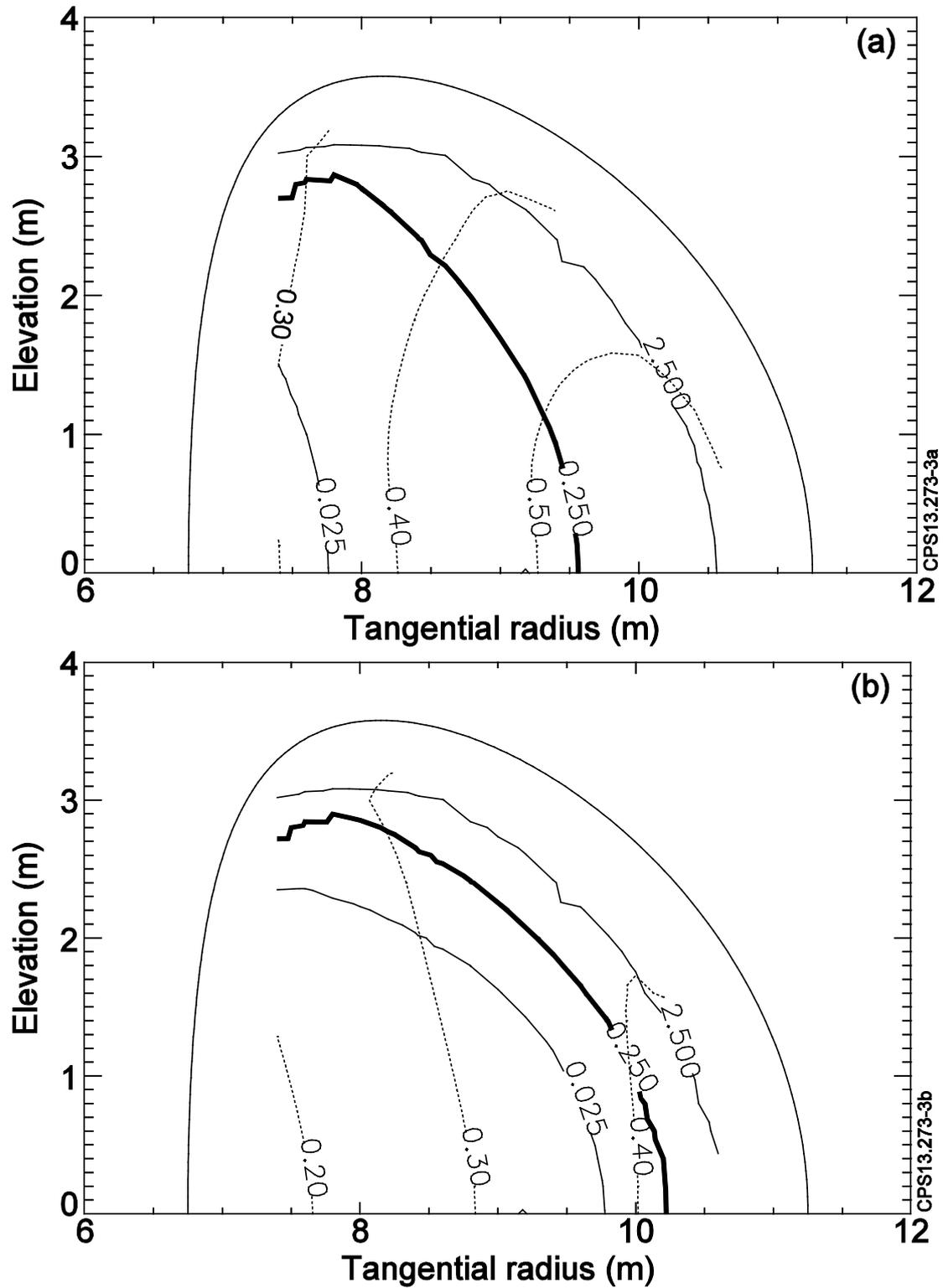

Fig.3. Contours of % shinethrough (solid lines) with $E_B$=1.5MeV for a) flat density and b) peaked density scenarios. The shinethrough=0.25% limit is indicated in bold. Selected γ contours are also shown (dashed lines).

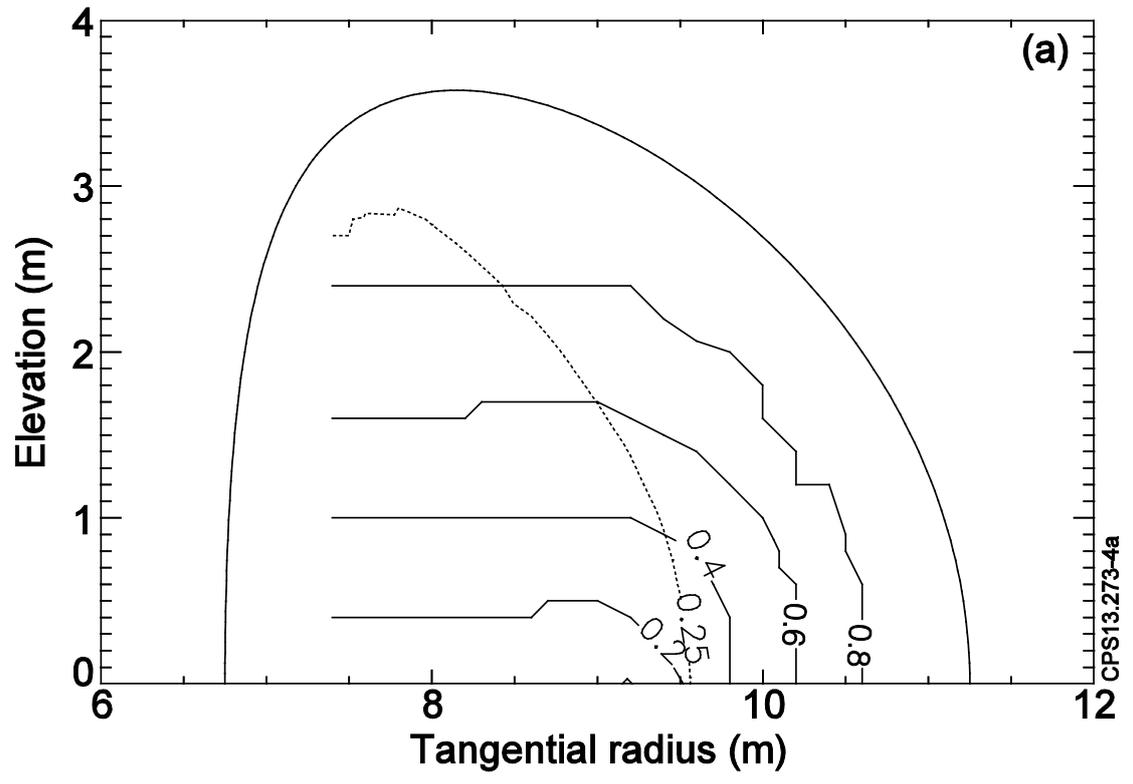

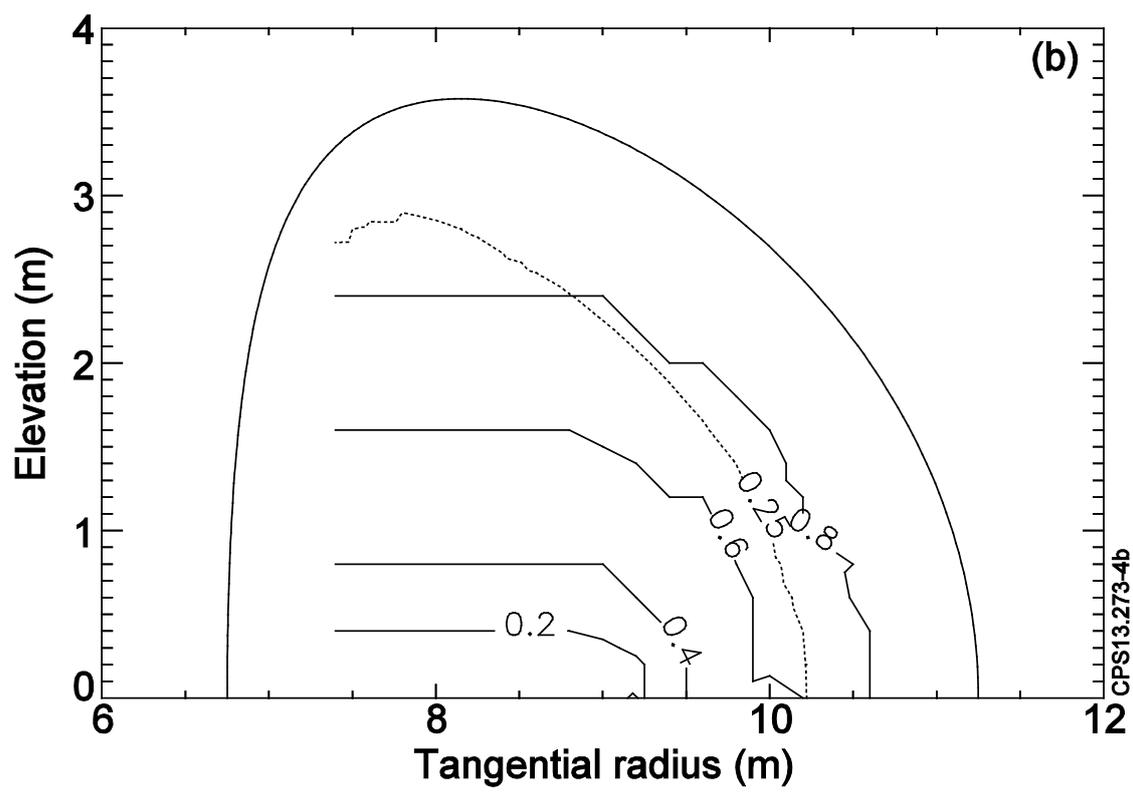

Fig.4. Contours showing the r/a location of the maxima of the NBCD profiles with $E_B=1.5MeV$ for a) flat density and b) peaked density scenarios. The shinethrough=0.25% limit is indicated (dashed).

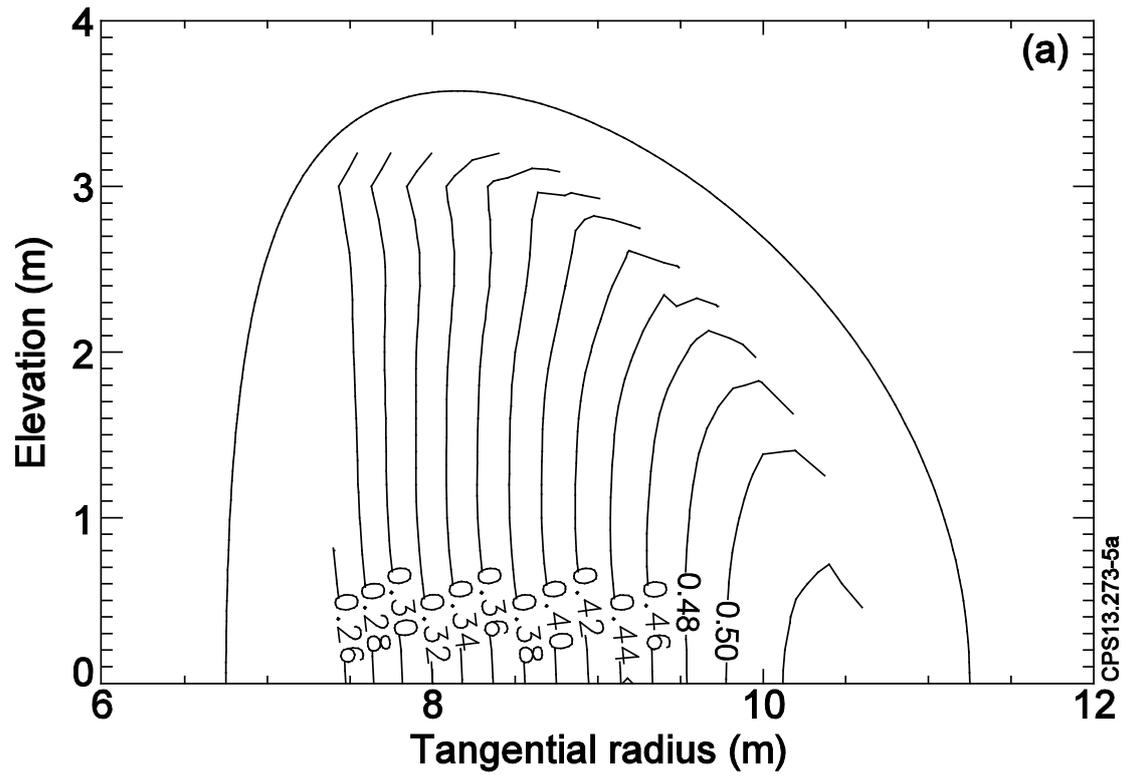

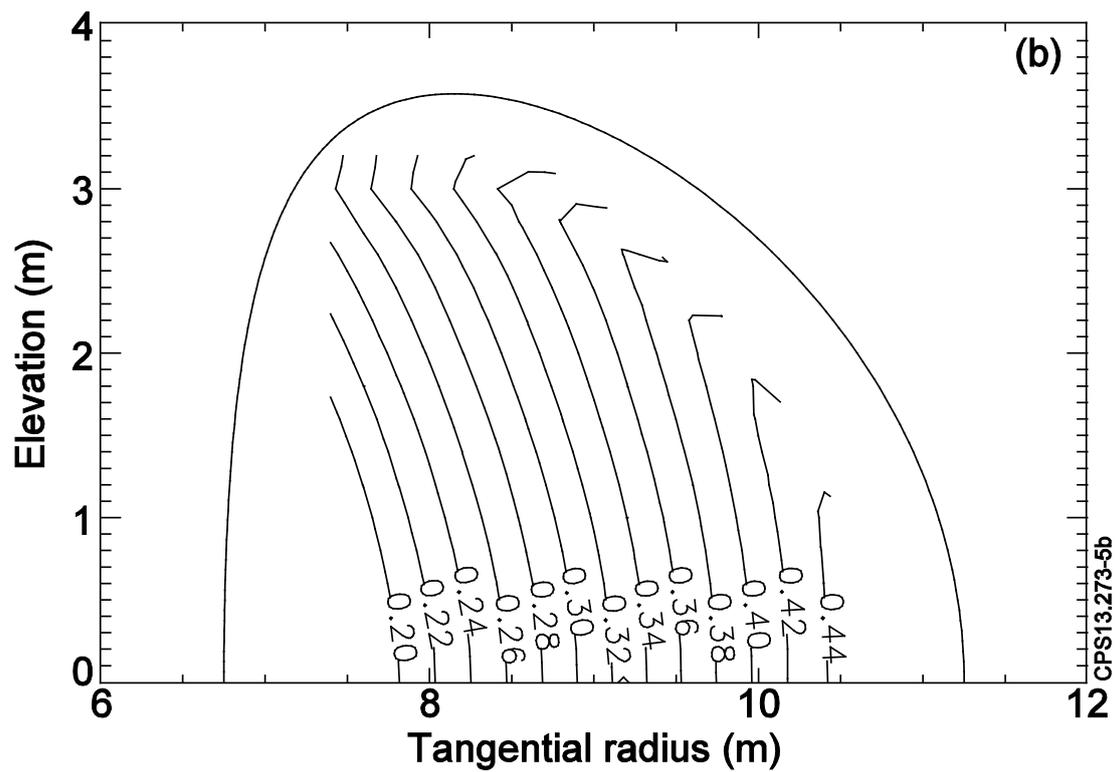

**Fig.5 Contours of NBCD efficiency, γ, with $E_B$=1.0MeV for a) the flat density scenario, b) the peaked density scenario.**

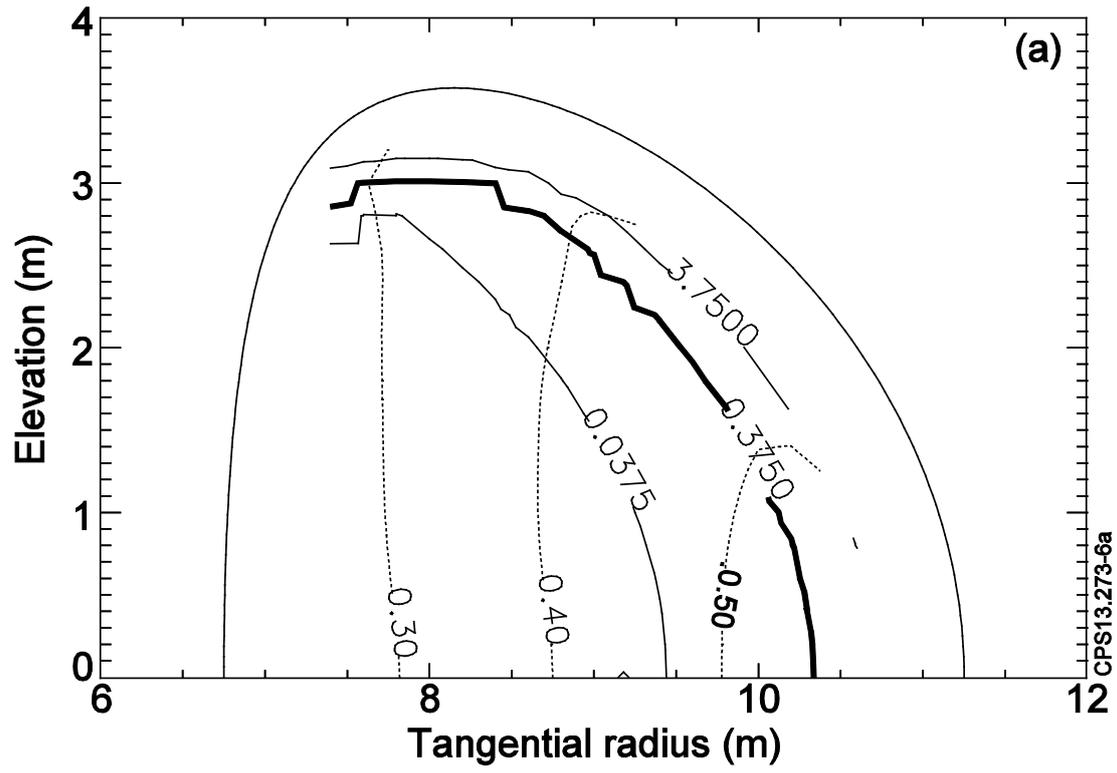

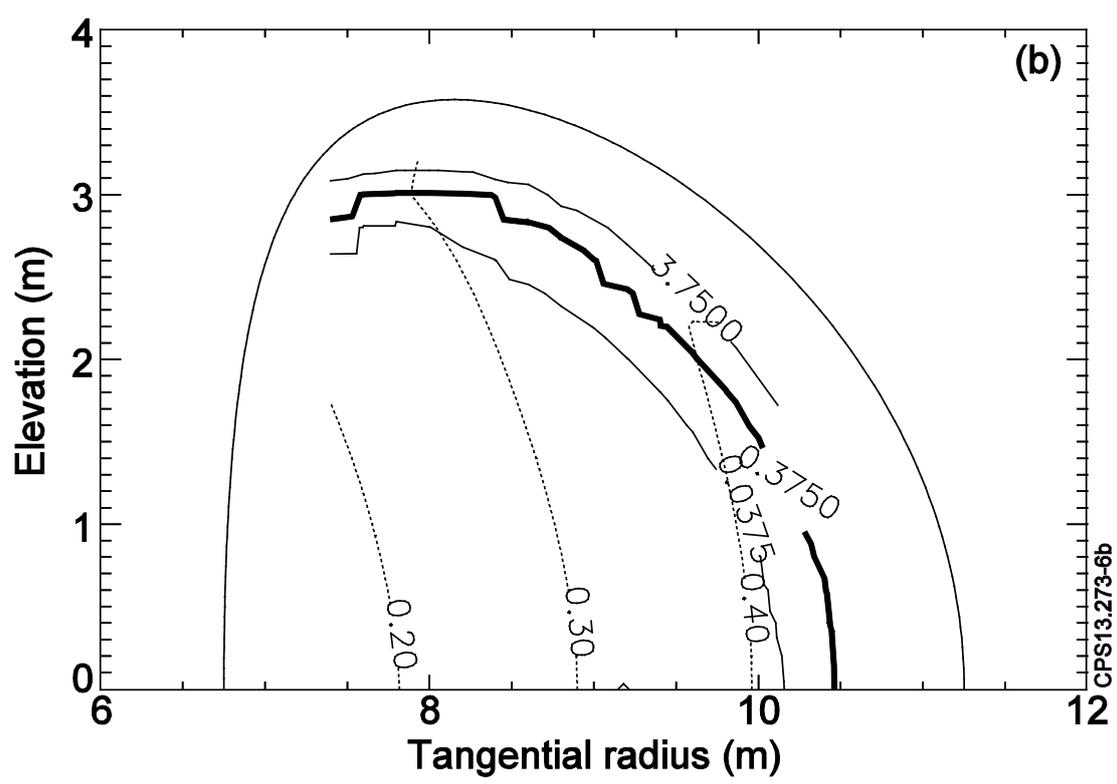

**Fig.6.** Contours of % shinethrough (solid lines) with $E_B=1.0$ MeV for a) flat density and b) peaked density scenarios. The shinethrough=0.375% limit is indicated in bold. Selected $\gamma$ contours are also shown (dashed lines).

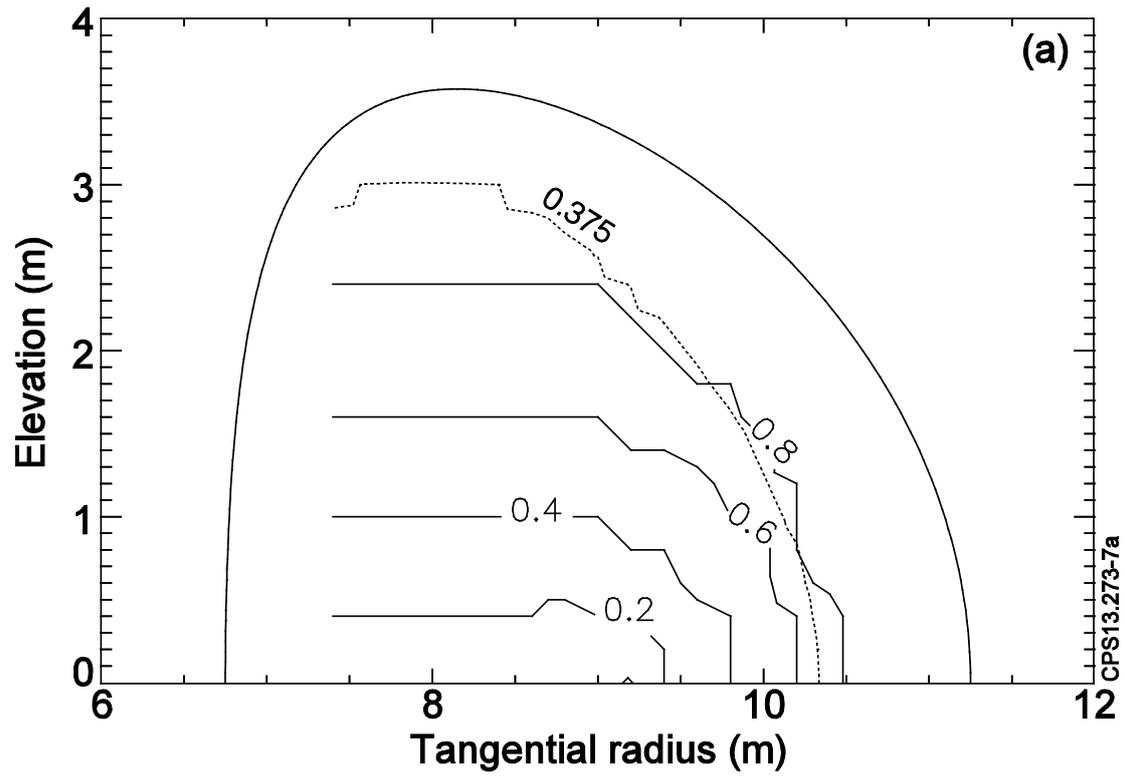

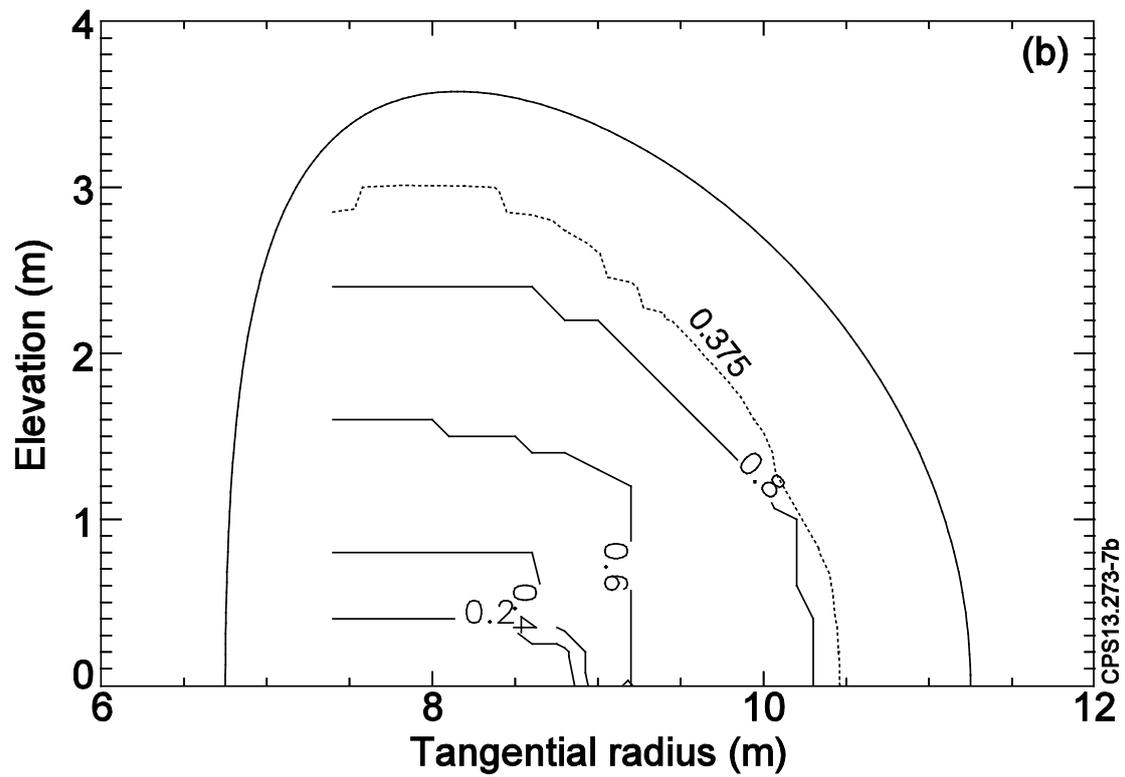

Fig.7. Contours showing the r/a location of the maxima of the NBCD profiles with $E_B=1.0$MeV for a) flat density and b) peaked density scenarios. The shinethrough=0.375% limit is indicated (dashed).

| Position | 1 | 2 | 3 | 4 | 5 | ... | n | crossover rate | mutation rate |
|---|---|---|---|---|---|---|---|---|---|
| Bit String 1 | 1 | 0 | 0 | 1 | 0 | . | 1 | 01001101 | 01000101 |
| Bit String 2 | 11001011 | 10001101 | 00101100 | 01100101 | 00100111 | . | 00110110 | 11100100 | 01000110 |

**Fig.8. A schematic representation of a trial solution with each bit string including a binary representation of the crossover and mutation rates associated with that particular trial solution and bit-string.**

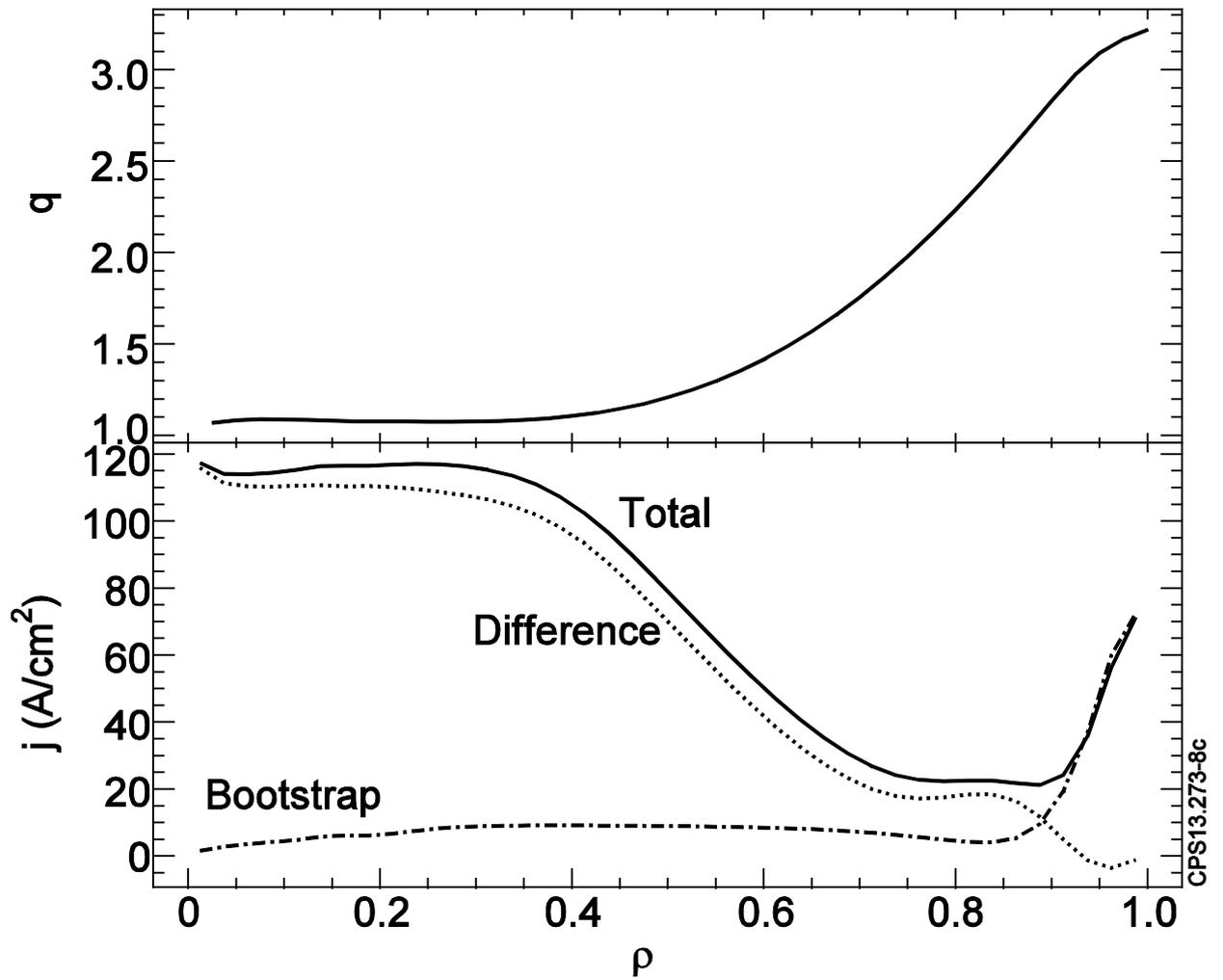

Fig. 9. Typical 'hybrid' q-profile (upper) and corresponding current profiles (lower). The dotted line is the difference between the total (solid line) and bootstrap current (dash-dotted), here calculated for the flat density scenario, and represents the current required to be provided by ohmic induction, NBCD, or a mixture of the two.

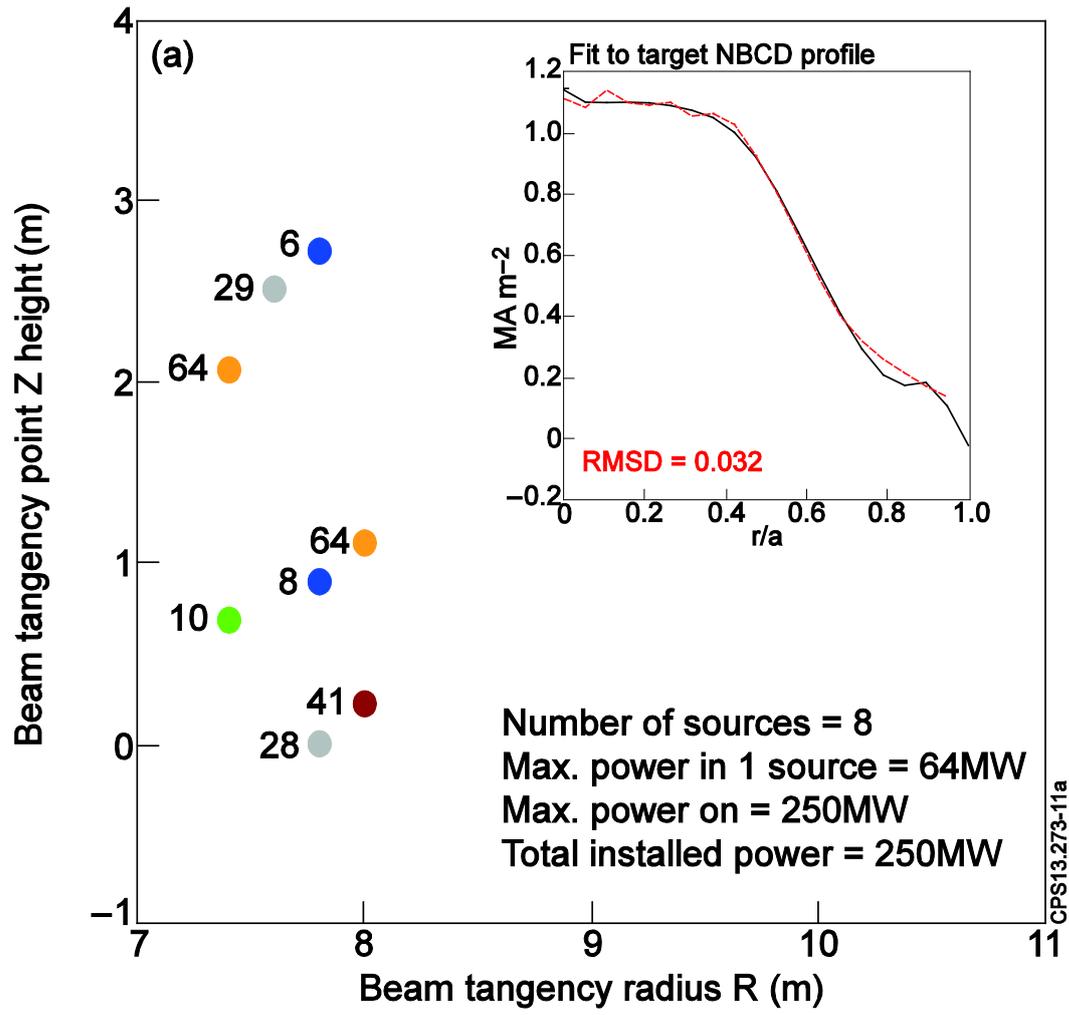

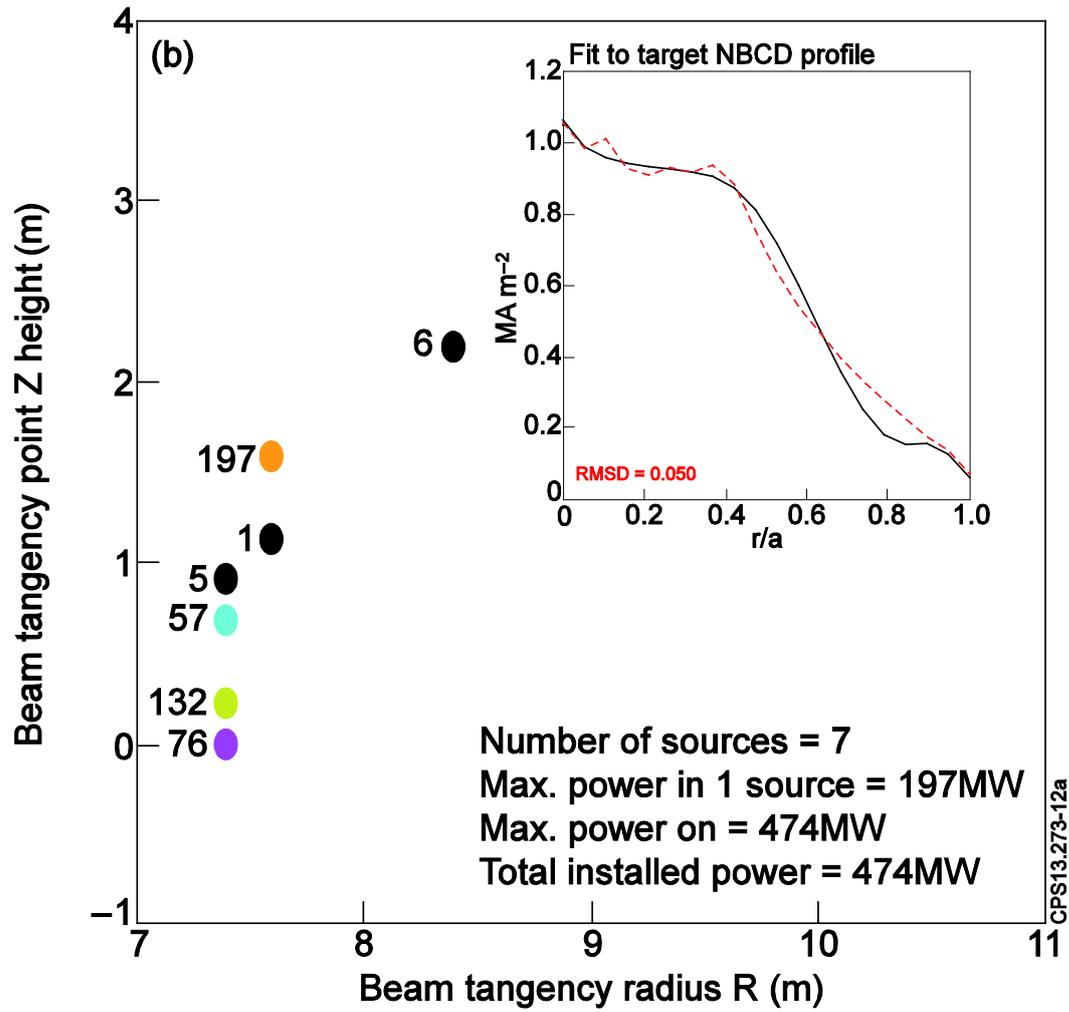

Fig. 10. Fits (dashed line, inset) after 1000 generations of genetic algorithm for the $E_B$=1.5MeV NBCD source functions to target j-profile (solid line, inset) and the distribution of power required for a) flat density, b) peaked density scenarios. The projection of each beam trajectory Z is at the beam tangency point R.

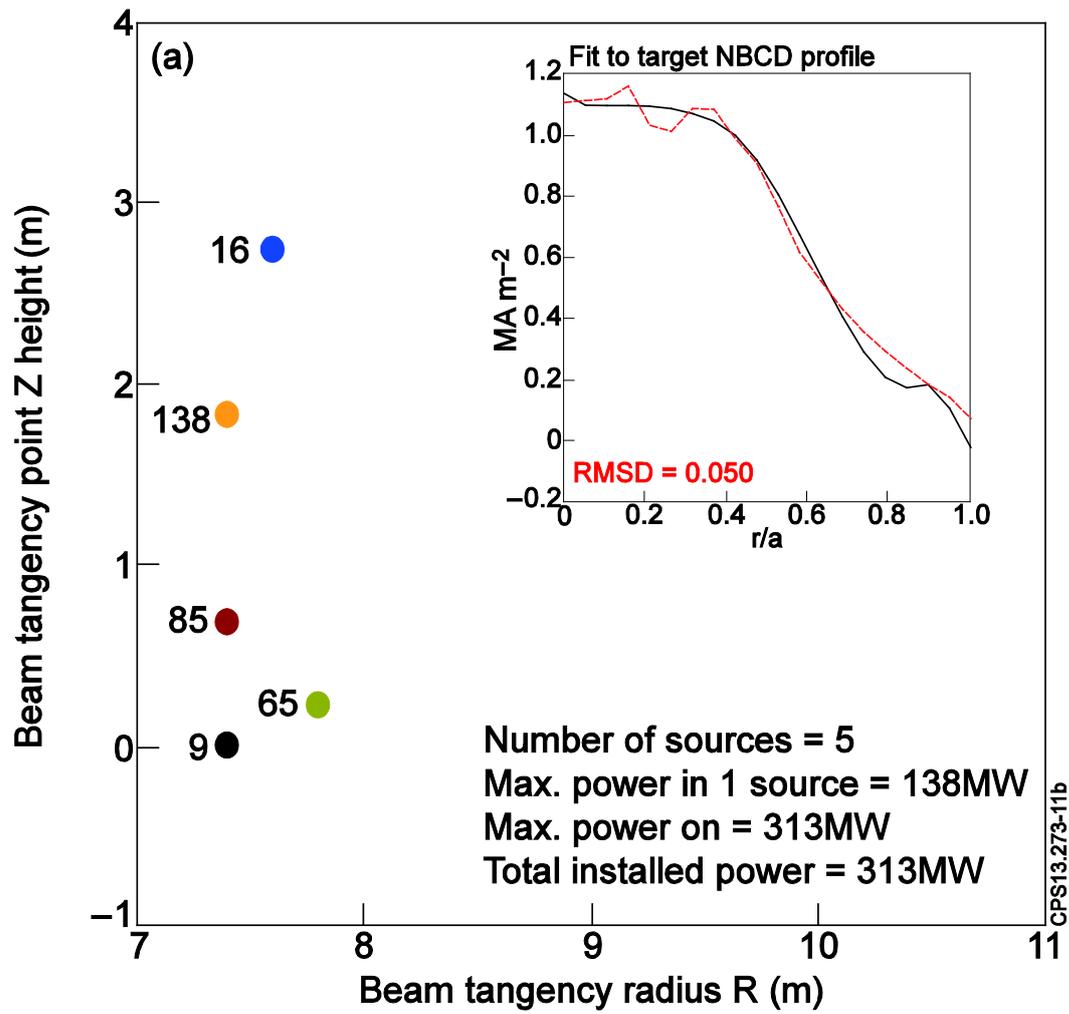

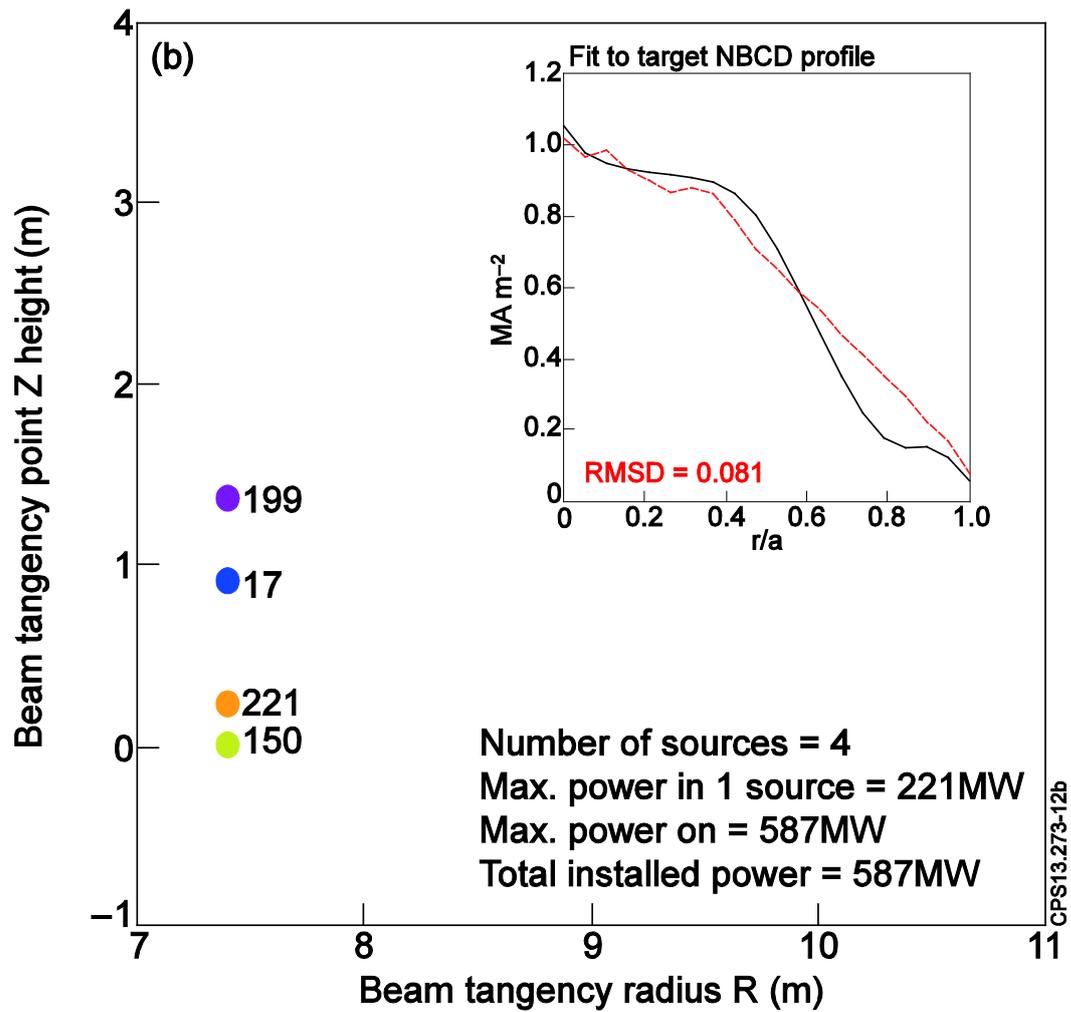

Fig. 11. Fits (dashed line, inset) after 1000 generations of genetic algorithm for the $E_B=1.0$MeV NBCD source functions to target j-profile (solid line, inset) and the distribution of power required for a) flat density, b) peaked density scenarios. The projection of each beam trajectory Z is at the beam tangency point R.